\documentclass{optica-article}

\journal{opticajournal}
\articletype{Letter}

\usepackage{lineno}
\usepackage{subcaption} 

\begin{document}

\title{Continuous Phase-Shifting Holography Utilizing a Moving Diffraction Grating}

\author{G.S. Kalenkov,\authormark{1,4,*} S.G. Kalenkov,\authormark{2} A.E. Shtanko,\authormark{3} B.C. Quirk,\authormark{4} and R.A. Mclaughlin\authormark{4}}

\address{\authormark{1}School of Mathematical and Physical Sciences, University of Technology Sydney, Ultimo, NSW 2007, Australia\\
\authormark{2}Optoelectronics Department, Moscow Polytechnic University, B. Semenovskaya 38, Moscow 107023, Russia\\
\authormark{3}Moscow State University of Technology “Stankin,” Vadkovskiy per., 1, Moscow 127994, Russia\\
\authormark{4}Faculty of Health and Medical Sciences, University of Adelaide, Adelaide SA, Australia}

\email{\authormark{*}george.kalenkov@uts.edu.au} 


\begin{abstract*} 
Fabrication of optical coherence tomography (OCT) fiber probes in cardiology involves sequentially splicing and cleaving multiple fibers to form a lens. During this process, the splice location-normally invisible under standard illumination-must be identified with micron-level accuracy. This paper presents an approach for splice detection using digital lensless microscopy, which is well-suited for deployment within the restricted space constraints typically found in fiber-fabrication setups. The optical setup incorporates a movable Ronchi grating that simultaneously acts as a beam splitter and a phase modulator. This design enables spatial separation of the reference and object beams while preserving the inherent stability of common-path interferometry. The acquired hologram set is processed using the continuous phase-shifting technique, which is known for its robustness against phase-shifting errors. The theoretical foundations for recording such holograms have been developed, and the conditions under which background noise and the conjugate order are suppressed have been identified. Experiments demonstrated the acquisition of high-contrast quantitative phase images of fiber splices with a spatial resolution down to 1 $\mu m$.
\end{abstract*}

\section{Introduction}
\begin{figure}[h!]
    \centering
    \begin{subfigure}{0.5\textwidth}
        \includegraphics[width=\linewidth]{ 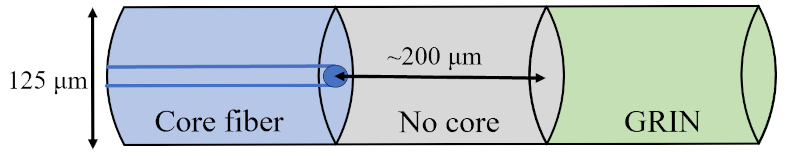}
        \caption{}
        \label{fig:probe-a}
    \end{subfigure}
    \begin{subfigure}{0.35\textwidth}
        \includegraphics[width=\linewidth]{ 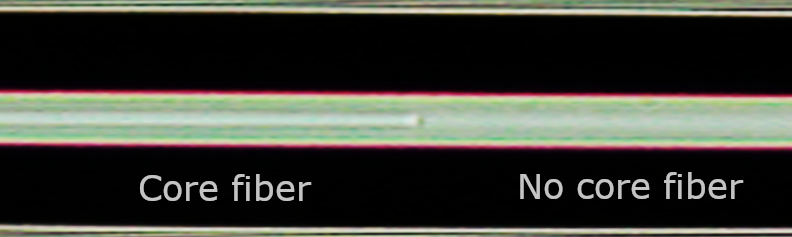}
        \caption{}
        \label{fig:probe-b}
    \end{subfigure}
    \caption{Fiber splice. a) A scheme showing two sequential splices between types of fiber. b) A microscope image of the slice between core and no core fibers.}
    \label{fig:probe}
\end{figure}

 Fiber-optic imaging probes are widely used in intravascular and endoscopic medical imaging. In particular, optical coherence tomography is the clinical gold-standard for imaging of atherosclerotic plaques in heart disease \cite{Araki2022}. A common method to fabricate the miniaturized lens required for these probes is to splice precise lengths of no-core and graded-index (GRIN) fiber to the distal end of a single mode fiber (SMF) \cite{Schmitt2015}. The optical properties of the lens depend critically on the lengths of the sections of no-core and GRIN fibre (Fig. \ref{fig:probe-a}.)
In practice, achieving micron-level length accuracy in fiber splicing involves attaching an extended length of no-core fiber to a single-mode fiber (SMF), identifying the splice, and cleaving at a precise distance from this point. This process is repeated to incorporate a GRIN fiber. The splice between the SMF and no-core fiber is very difficult to detect using conventional microscopy due to the optical similarities of the fibers, differentiated only by the small core of the SMF. Current techniques to visualize the discontinuity of the splice, as shown in Fig. \ref{fig:probe-b}, require precise focusing and high numerical aperture optics enforcing a short working distance that is impractical in in-line fiber-optic fabrication processes. This presents significant challenges in the precise fabrication of these fiber-optic probes. Lensless digital holographic microscopy (i.e. no lens between sample and sensor) offers a promising solution for identifying fiber inhomogeneities \cite{001_1450338}. This technique records the interference pattern generated by the interaction of two waves—one of which is modulated by the object—on a detector (such as a CCD or CMOS sensor), followed by computational reconstruction of the object’s wave field amplitude and phase.
In the context of a commercial manufacturing environment, the optical system for hologram recording must be robust against vibrations and external noise. The Ronchi grating interferometer, a variant of the common-path interferometer, is particularly suited to meet these requirements. Ronchi grating interferometers can function as either shearing \cite{002_Wu:19} or holographic interferometers \cite{003_Popescu:06,004_Zeev:09,005_MA2020124563,006_Tsai:24,007_Wang,008_jeong2021common}. In holographic configurations, the diffraction grating serves as a beam splitter, dividing the incoming wave into multiple diffraction orders, with two of these orders generating the interference pattern on the detector. In such systems \cite{003_Popescu:06,004_Zeev:09,005_MA2020124563,006_Tsai:24,007_Wang,008_jeong2021common}, the wave field passing through the object and objective lens is split into two orders by the grating. Object information is removed from one of these orders through spatial filtering, allowing it to function as a reference wave. The grating thus serves a dual purpose, acting both as a beam splitter \cite{003_Popescu:06,007_Wang} and a phase modulator for the reference wave \cite{004_Zeev:09,005_MA2020124563,006_Tsai:24,008_jeong2021common}. Depending on its position relative to the Fourier plane of the objective lens, holograms may be recorded with either an on-axis \cite{004_Zeev:09} or off-axis \cite{005_MA2020124563,006_Tsai:24,007_Wang,008_jeong2021common} reference beam. The phase image reconstructed from these holograms is highly informative. However, this approach poses challenges related to the suppression of the twin-image and zero-order diffraction background. Solutions include angular spectrum filtering algorithms \cite{009_Takeda:82} and phase-stepping techniques \cite{010_Yamaguchi:97}, the latter of which involves sequentially capturing interference patterns while incrementally adjusting the phase of the reference wave. In Ronchi grating-based systems, these phase adjustments are typically achieved by shifting the grating transversely by a fraction of its period \cite{004_Zeev:09,005_MA2020124563,006_Tsai:24,007_Wang,008_jeong2021common}.
In this work, we expand the capabilities of a Ronchi grating-based holographic interferometer by focusing on the registration of Fresnel holograms of the object, providing new insights and potential advantages compared to image plane holography. In this approach, each element of the object is not confined to a specific detector pixel but instead its diffraction pattern is recorded over the entire detector surface, thereby reducing noise. Additionally, we place the object not before the grating but after it, limiting its presence to only one of the diffraction orders. This configuration allows the unmodulated diffraction order to function as a spherical reference beam, eliminating the need for spatial filtering to purify the reference beam—an essential step in previous studies.
To record the holograms and reconstruct amplitude and phase images, we employed a continuous phase-shifting technique \cite{011_Kalenkov:20}, where the phase of the reference wave is linearly varied over time. Multi-frame hologram acquisition was performed at regular time intervals, enabling each pixel to capture an interferogram as a function of time. By applying a Fourier transform to the time-domain data and isolating the component with the maximum amplitude, we determined the complex amplitude of the wave field at each pixel. Extending this process across the entire detector yields a two-dimensional field distribution of the object. This method significantly enhances image quality. Furthermore, the absence of an objective lens between the fiber sample and the sensor allows for direct access to the object, facilitating experimental interventions and reducing space constraints when integrated into an industrial fabrication process. The effectiveness of this approach is demonstrated through the visualization of material inhomogeneities in optical fiber. 

\section{Theory}
\begin{figure}[ht]
\centering
\includegraphics[width=250px]{ 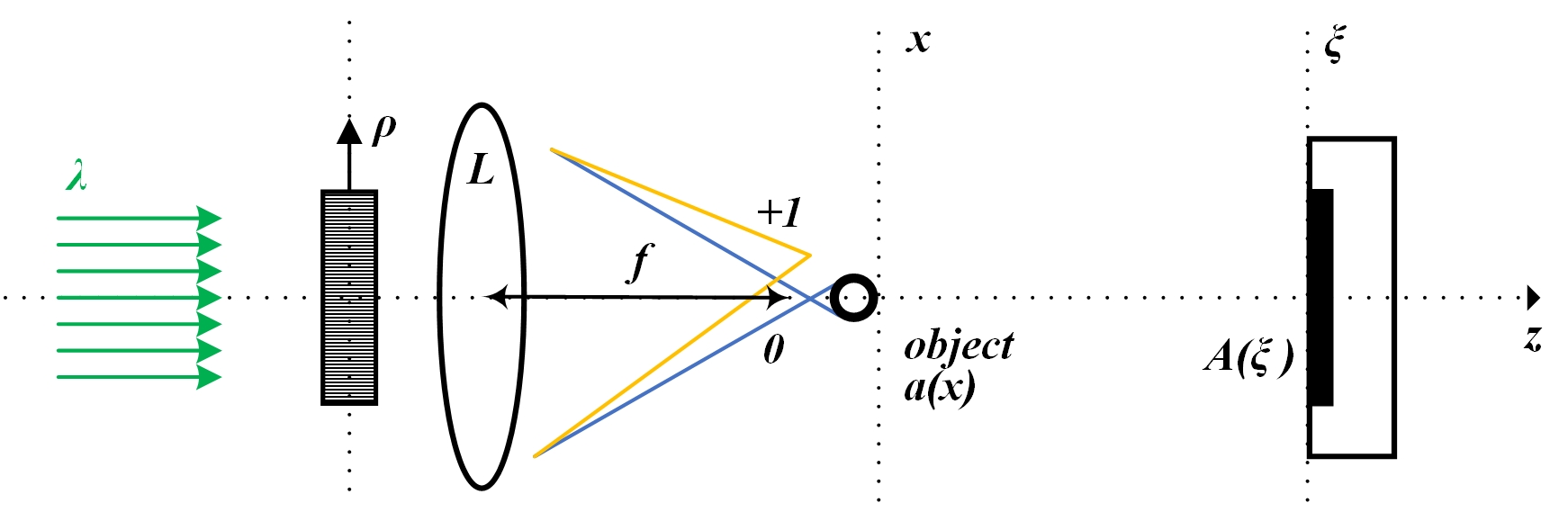}
\caption{Principle optical scheme. $\lambda$ -- the wavelength, $\rho$ -- vector of the direction of grating displacement,  L -- Lens or an objective,  f -- focal distance,  $\boldsymbol{x}$ -- an object plane, $\boldsymbol{\xi}$ -- the plane of registration,diffraction grating. }
\label{fig:theory_scheme}
\end{figure}

Let a plane monochromatic wave of unit amplitude with the wavelength $\lambda$ fall normally onto a grating Fig.~\ref{fig:theory_scheme}. In the focal or Fourier plane of lens $L$, the $0^{th}$ and $1^{st}$ diffraction orders of the light form an object and a reference  source, respectively. A transparent object is placed at a distance from the $0^{th}$ order point source in the diverging beam. This distance is experimentally determined to ensure that the light from the point source of the zero-order diffraction fully illuminates the object. Denote $a(\boldsymbol{x})$ as the complex amplitude of the wavefield immediately behind the object, and $A(\boldsymbol{\xi})$ as the complex amplitude of the diffracted field in the registration plane, where $\boldsymbol{x}$ and $\boldsymbol{\xi}$ are the respective coordinates in the object and registration planes. These fields are linked by an integral transformation: $A(\boldsymbol{\xi}) = \hat{\Phi}[a(\boldsymbol{x})],$ where $\hat{\Phi}$ denotes the operator transforming the field from the plane $\boldsymbol{x}$ to the registration plane $\boldsymbol{\xi}$. Depending on the geometry, the operator $\hat{\Phi}$ represents either a Fourier transform or a Fresnel transform. 
When the diffraction grating of period $d$ is displaced transversely to the optical axis $z$ by a distance $\rho$, the reference wave acquires a phase shift given by
$\varphi(\rho) = 2\pi u \rho$, where $u = 1/d$ is the spatial Fourier frequency corresponding to the $1^{st}$ order. The dependence of the complex amplitude of the reference wave $r(\rho)$ on the displacement $\rho$ is:
$r(\rho) = r \exp ( -2\pi i u \rho) = r \exp ( -2\pi i \rho /d)$. The intensity of the interference field $I(\boldsymbol{\xi},\rho)$ in the plane of registration is: 
\begin{equation}
I(\boldsymbol{\xi},\rho) = |{r + A}|^2 = I_{bg} + A(\boldsymbol{\xi})r^*\exp (2\pi i \rho /d) + A(\boldsymbol{\xi})^*r \exp (-2\pi i \rho /d),
\label{eq:1}
\end{equation}
where $I_{bg} = |{A}|^2+|{r}|^2$ is the background, the symbol * denotes complex conjugation. In Eq.~(\ref{eq:1}) we have neglected mutual cross-coherence function of the fields, assuming it to be unity. Also, all insignificant factors have been omitted. We now multiply Eq.~(\ref{eq:1})
by $\exp (-2\pi i \rho /d)$ and integrate over the specified limits $-L_m /2 \leq \rho \leq L_m /2$. Considering that the functions $I_{bg}$, $A$, and $A^*$ do not depend on the variable $\rho$, from Eq.~(\ref{eq:1}) we obtain:
\begin{equation}
\begin{array}{l}
\int_{-L_m/2}^{L_m/2} I(\boldsymbol{\xi},\rho) exp(2\pi i\rho/d)  d\rho = \\
= A(\boldsymbol{\xi}) r^* + I_{bg} \int_{-L_m/2}^{L_m/2} \exp(-2\pi i \rho / d) d\rho + A^*(\boldsymbol{\xi}) r \int_{-L_m/2}^{L_m/2} \exp(-4\pi i \rho / d) d\rho.
\end{array}
\label{eq:2}
\end{equation}
We require that the integrals appearing in Eq.~\eqref{eq:2} vanish.
\begin{equation}
\int_{-L_m/2}^{L_m/2}{exp{(}-2\pi i\rho/d)d\rho}=L_m\mathrm{sinc}\left(L_m/d\right)\mathrm{=0}
\label{eq:3}
\end{equation}
\begin{equation}
\int_{-L_m/2}^{L_m/2}{exp{(}-4\pi i\rho/d)d\rho}=L_m\mathrm{sinc}(2L_m/d)\mathrm{=0}
\label{eq:4}
\end{equation}
The requirement that the integrals in Eq.~\eqref{eq:3} and Eq.~\eqref{eq:4} vanish imposes an obvious condition on the limits of integration: $L_m=md$. Thus, in each pixel $\boldsymbol{\xi}$, it is necessary to calculate the integral of the interferogram over symmetric limits such that the integration interval equals an integer number of grating periods. Under this condition Eq.~\eqref{eq:5} yields the complex amplitude of the object in the plane of registration:
\begin{equation}
    A(\boldsymbol{\xi})=\frac{1}{r}\int_{-L_m/2}^{L_m/2}{I(\boldsymbol{\xi},\rho)exp{(}2\pi i\rho/d)}d\rho
    \label{eq:5}
\end{equation}

\section{Experiment}
\begin{figure}[h]
\centering
\includegraphics[width=\linewidth]{ 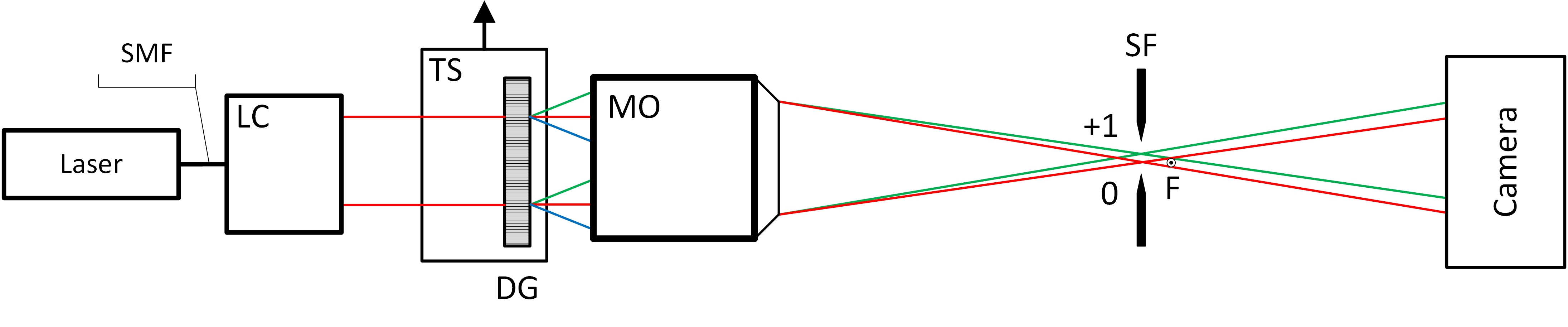}
\caption{Experimental setup. SMF - single mode fiber, LC -- Laser collimator, TS -- translation stage, DG -- diffraction grating, MO -- micro-objective, SF -- spatial filter, F -- fiber.}
\label{fig:scheme_exp}
\end{figure}

\begin{figure}[h!]
    \centering
    \begin{subfigure}{0.49\textwidth}
        \includegraphics[width=\linewidth]{ 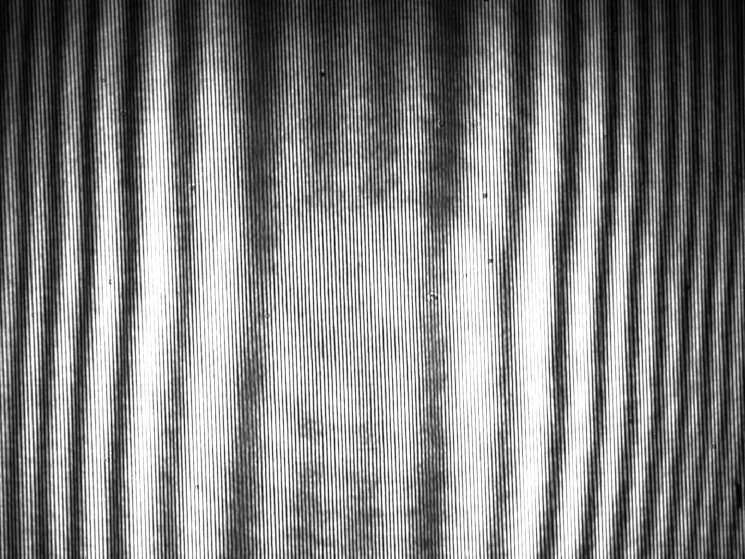}
        \caption{}
        \label{fig:rawdata-a}
    \end{subfigure}
    \begin{subfigure}{0.48\textwidth}
        \includegraphics[width=\linewidth]{ 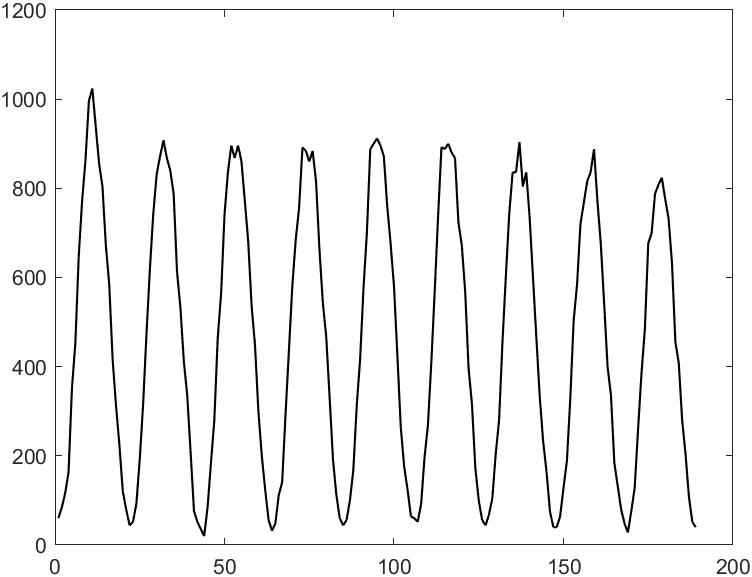}
        \caption{}
        \label{fig:rawdata-b}
    \end{subfigure}
    \caption{Registered data. a) One of the holograms acquired during the registration process. b) An interferogram $I(\rho)$ registered in a single pixel $\boldsymbol{\xi}$ = (300,300), satisfying the condition $L_m=md$. }
    \label{fig:rawdata}
\end{figure}

\begin{figure}[h!]
    \centering
    \begin{subfigure}{0.31\textwidth}
        \includegraphics[width=\linewidth]{ 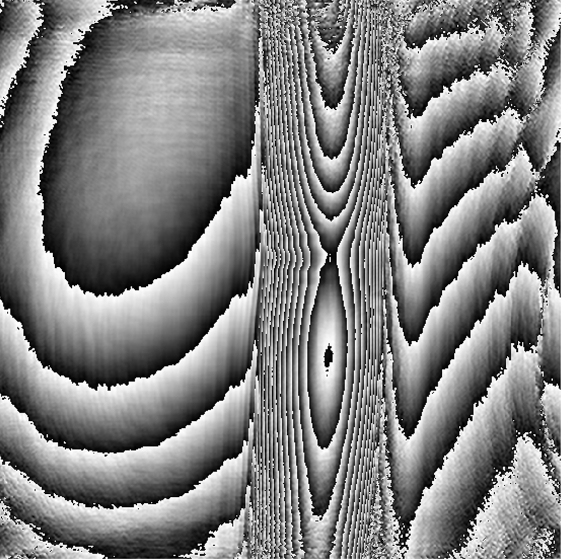}
        \caption{}
        \label{fig:phase-a}
    \end{subfigure}    
    \begin{subfigure}{0.15\textwidth}
        \includegraphics[width=\linewidth]{ 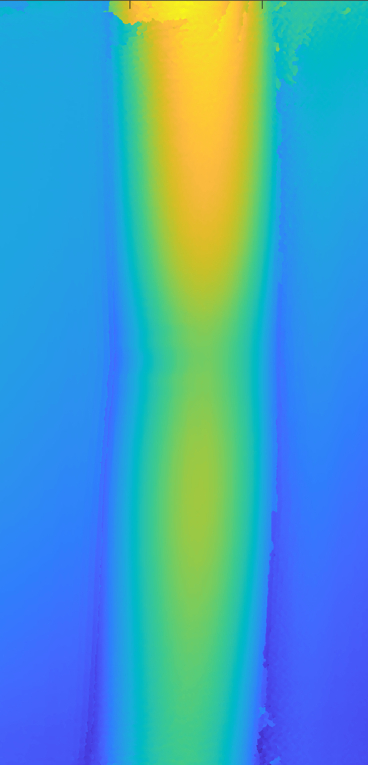}
        \caption{}
        \label{fig:phase-b}
    \end{subfigure}    
    \begin{subfigure}{0.51\textwidth}
        \includegraphics[width=\linewidth]{ 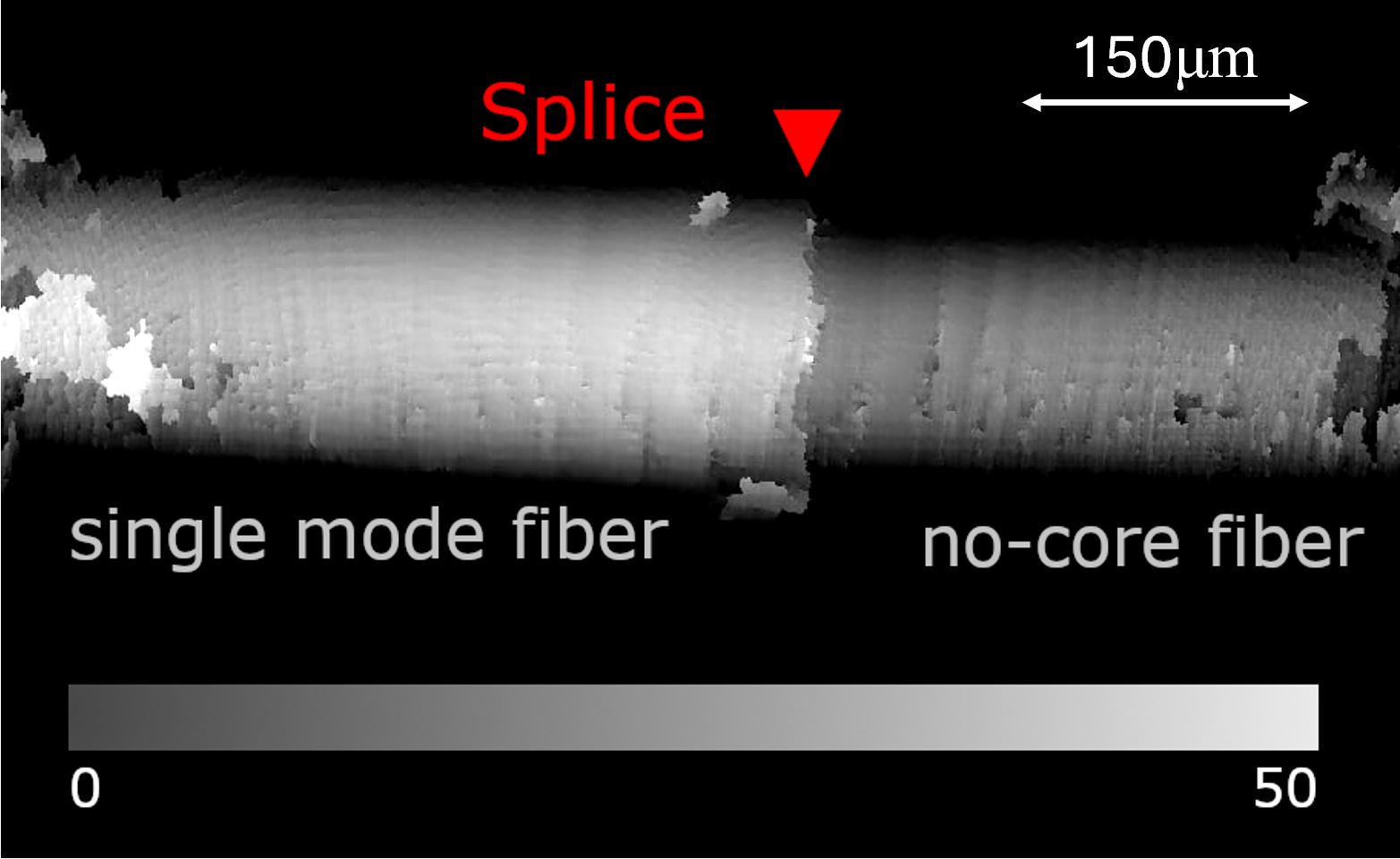}
        \caption{}
        \label{fig:phase-c}
    \end{subfigure}
    \caption{Phase images of the single mode/no-core fiber splice in radians. a) Wrapped phase image $\varphi = [-\pi/2\;\pi/2]$. b) Unwrapped phase image $\varphi = [357\;3482]$. c) Unwrapped phase image of the fiber splice after removing a spherical phase carrier and limiting the phase range ($\varphi = [1539\;1589]$) to 50 radians, highlighting the significant part with high contrast.}
    \label{fig:phase}
\end{figure}

The optical scheme of the experimental setup is depicted in Fig. \ref{fig:scheme_exp}. Collimated laser light ($\lambda = 640nm$) is directed onto a transmission diffraction grating (40 lines per mm) mounted on a translation stage (Model X-LSM100A, Zaber, Canada). The light passes through a microobjective ($10 x, 0.45 NA$), with an aperture mask placed in its focal plane, allowing zero and first-order diffraction to pass while blocking others. The fiber is illuminated by the zero-order beam, forming the object wave, while the first-order beam serves as the reference wave.
The interference intensity of these two waves Eq.~\eqref{eq:1}, a digital hologram, is recorded by a sensor array. A series of digital holograms is captured while the grating is moved perpendicular to the beam (see \textbf{Visualization 1}). One of such holograms is shown in Fig.~\ref{fig:rawdata-a}. Hence, each pixel of the sensor registers an interferogram (Fig.~\ref{fig:rawdata-b}) which is processed according to Eq.~\eqref{eq:5} and allows for the computation of the object's complex field $A(\boldsymbol{\xi})$ in each pixel in the sensor plane \cite{011_Kalenkov:20}. 
Then, a zero-order alignment is applied to all spectra $A(\boldsymbol{\xi})$ to reduce the spatial frequency. Fresnel transformation is then used to calculate the complex field in the object plane $a(\boldsymbol{x},z=z_{obj})$. The phase of the object $\varphi(\boldsymbol{x})=imag[a(\boldsymbol{x})]/real[a(\boldsymbol{x})]$ is calculated Fig.~\ref{fig:phase-a}. To correct for the spherical wavefront carrier produced by the single point illumination source, a one-time calibration procedure has to be performed. The illumination field $A_{ill}(\boldsymbol{\xi})$ is registered without the object in the beam path. It is also back-propagated to the object's plane $a_{ill}(\boldsymbol{x},z=z_{obj})$. Both fields $a(\boldsymbol{x},z=z_{obj})$ and $a_{ill}(\boldsymbol{x},z=z_{obj})$ are phase-unwrapped \cite{012_Herraez:02}. The subtraction $a(\boldsymbol{x}) - a_{ill}(\boldsymbol{x})$ eliminates the spherical carrier and flattens the background. The final result is presented on Fig.~\ref{fig:phase-c}. The typical 'phase jump' from SMF to no-core fiber is of an order of 10 radians. The field of view measures $0.5 mm$ across 1003 pixels. With the numerical aperture (NA) of 0.45, the resulting resolution is approximately $1.11\lambda$ or $0.7 \mu m$. 

\section{Discussion}
It should be noted that Eq.~\eqref{eq:5}, which describes the procedure for computing the complex field of an object recorded using a movable grating $A(\boldsymbol{\xi}, \rho)$, is identical to the corresponding equation for $A(\boldsymbol{\xi}, \delta)$ derived in the continuous phase-shifting method, where the mirror in the reference arm of the interferometer moves at a constant velocity \cite{011_Kalenkov:20, 013_Kalenkov:21}. 

The proposed optical setup is particularly well-suited for detecting fiber splices due to the strong dispersive properties of optical fibers. An intensity of approximately $6\%$ of the total beam power is directed into the first order of diffraction by the Ronchi grating provides a balanced ratio between the reference and object beams, leading to optimal conditions for registration.

The proposed configuration combines the advantages of a common-path interferometer, where the grating functions as both a beam splitter and a modulator, ensuring high vibration resistance and reducing the requirements for temporal coherence and stability of the laser source. On the other hand, this configuration spatially separates the object and reference beams, eliminating the need to filter the reference beam from the object field. The Ronchi grating provides significant beam separation, allowing the fiber to be positioned in one order without overlapping the others. At the same time, the small distance between the sources lowers the carrier spatial frequency on the detector.

The accuracy requirements for grating movement are lower compared to phase-shifting methods based on mirror-scanning approaches, when the mirror is displaced to produce a series of phase-delayed references. The reference beam is expressed as $r_1 = \exp[i\varphi_1(z)] = \exp\left(\frac{2\pi i}{\lambda} z\right)$, where $z$ is the displacement of the mirror. For a movable grating, the reference beam undergoes a different phase modulation due to the displacement $z$ across the beam and is expressed as $r_2 = \exp[i\varphi_2(z)] = \exp\left(\frac{2\pi i}{d} z\right)$. For the same displacement error of the moving element (whether it is the mirror or the grating), the phase error in the first case is $\Delta\varphi_1 = \frac{2\pi \Delta z}{\lambda}$, while in the second case it is $\Delta\varphi_2 = \frac{2\pi \Delta z}{d}$. Given that $\lambda / d \ll 1$, it is evident that the accuracy requirements for grating movement are less stringent. Furthermore, in our setup these requirements are relaxed even more due to the use of the continuous phase-shifting technique, which is more robust against phase modulation errors in the reference beam compared to 3- or 4-step phase-shifting techniques \cite{011_Kalenkov:20}.

An additional advantage is the simplicity of the setup, which consists of a translation stage, a grating, a standard objective, and a camera without additional optics.

We note that the goal was not to precisely determine the thickness or refractive index of the fibers. The objective was to accurately locate the fiber splice, as this is the critical parameter in accurately cleaving fiber lengths in an industrial setting to achieve reliable optical parameters of NFC/GRIN fiber-optic imaging probes. Therefore, the absolute phase values are not critical (Fig.~\ref{fig:phase-b}); instead, the relative values defining the contrast between the phase images of core and no-core fibers is of interest.

\section{Conclusion}
An optical design for a lensless digital holographic microscope employing a movable Ronchi grating and a continuous phase-shifting method is proposed. Experiments were carried out to demonstrate the acquisition of high-contrast quantitative phase images of fiber splices with spatial resolution down to 1 µm, facilitating the automation of splice detection in fiber-fabrication setups.

\section*{Disclosures} The authors declare no conflicts of interest.

\bibliography{main}






\end{document}